\documentclass[a4paper,fleqn]{cas-sc}

\usepackage[numbers,sort&compress]{natbib}

\usepackage{subcaption}  

\def\tsc#1{\csdef{#1}{\textsc{\lowercase{#1}}\xspace}}
\tsc{WGM}
\tsc{QE}
\tsc{EP}
\tsc{PMS}
\tsc{BEC}
\tsc{DE}

\begin{document}
\let\WriteBookmarks\relax
\def\floatpagepagefraction{1}
\def\textpagefraction{.001}

\shorttitle{Supervised cooperation on interdependent public goods games}

\shortauthors{Ling et~al.}

\title [mode = title]{Supervised cooperation on interdependent public goods games}                      



%
\author[1]{Ting Ling}
\affiliation[1]{organization={College of Artificial Intelligence, Southwest University},
    city={Chongqing},
    postcode={400715}, 
    country={China}}

\author[1]{Zhang Li}
\author[1]{Minyu Feng}
\cormark[1]
\ead{myfeng@swu.edu.cn}

\cortext[cor1]{Minyu Feng}

\author[2]{Attila Szolnoki}
\affiliation[2]{Institute of Technical Physics and Materials Science, Centre for Energy Research, P.O. Box 49, H-1525 Budapest, Hungary.}
\ead{szolnoki.attila@ek-cer.hu}

\begin{abstract}
It is a challenging task to reach global cooperation among self-interested agents, which often requires sophisticated design or usage of incentives. For example, we may apply supervisors or referees who are able to detect and punish selfishness. As a response, defectors may offer bribes for corrupt referees to remain hidden, hence generating a new conflict among supervisors. By using the interdependent network approach, we model the key element of the coevolution between strategy and judgment. In a game layer, agents play public goods game by using one of the two major strategies of a social dilemma. In a monitoring layer, supervisors follow the strategy change and may alter the income of competitors. Fair referees punish defectors while corrupt referees remain silent for a bribe. Importantly, there is a learning process not only among players but also among referees. Our results suggest that large fines and bribes boost the emergence of cooperation by significantly reducing the phase transition threshold between the pure defection state and the mixed solution where competing strategies coexist. Interestingly, the presence of bribes could be as harmful for defectors as the usage of harsh fines. The explanation of this system behavior is based on a strong correlation between cooperators and fair referees, which is cemented via overlapping clusters in both layers. 
\end{abstract}

\maketitle

\section{Introduction}\label{sec1}

Cooperation typically refers to the pro-social behavior in which individuals or groups work together and support each other in order to achieve their common goals \cite{1}. It involves interdependence and reciprocal relationships among different individuals which is widespread in nature and human society~\cite{perc_pr17}. According to Darwin's theory of evolution \cite{2}, cooperation is unsustainable, as individuals usually show selfish tendencies, and defection seems to be a more beneficial individual choice. However, in the real world, there are many instances of cooperative behavior~\cite{nowak_11,capraro2021mathematical}. For example, worker ants will voluntarily sacrifice themselves to protect the queen and the nest, and birds will form groups during migration to navigate and protect the weaker members of the group together. Thus, it has attracted much attention from researchers to find the explanation how cooperation arises spontaneously among selfish individuals \cite{4,liu_jq_csf24,6,wang_p_csf24,li2021evolution,li2024asymmetrical}. As the adequate theoretical approach, evolutionary game theory offers an effective and powerful framework to crack this enigma. It employs various game models, including Prisoner's Dilemma Game (PDG) \cite{rapoport_70, 11}, Snowdrift Game (SDG) \cite{10,li_k_csf21}, and Public Goods Game (PGG) \cite{perc_jrsi13}, to describe different social dilemma situations and quantify the relations of individuals' choices. PGG represents a typical multiplayer social dilemma where cooperators can choose to put a certain amount of resources to a public pool, and the enhanced contributions will be distributed equally among all individuals. Evidently, defectors can considered as ``free-riders'' who also benefit from the public pool without any investment. 

Over the years several cooperation supporting mechanisms were identified, including the ``main five'', as kin selection, direct reciprocity, indirect reciprocity, group selection and network reciprocity, summarized by Nowak in his seminal essay~\cite{21}. In addition, as a result of continuous development of evolutionary game theory, other positive aspects were also recognized such as memory effects \cite{22,danku_srep19,xu_zj_c19}, multi-gaming \cite{25,duong_jmb16}, reputation \cite{29,30,xia2023reputation}, and the usage of incentives, like rewards \cite{szolnoki_epl10,yan_ry_pla24,duong_prsa21} or punishment \cite{27,szolnoki_pre11b,zhang_y_csf24,cimpeanu_kbs21}.
Among all of these mechanisms, punishment is one of the most widely studied mechanisms~\cite{gao_sp_amc25,lee_hw_amc22,kang_hw_pla24,ohdaira_plr23,hua_sj_csf3}. In previous studies, punishment mechanisms are mainly divided into individual or peer punishment and collective or pool punishment. In the case of peer punishment, a punisher player faces with defectors directly and bears the cost of punishment proportionally to the defection level. In the alternative case, punishers leave this task to an institution. Accordingly, they have a constant extra cost to cover this institution independently of the presence of defectors.
Herrmann {\it et~al.} emphasized that punishment is beneficial in maintaining cooperation only when the society develops adequate norms~\cite{31}. Song {\it et~al.} proposed a conditionally neutral punishment mechanism, whereby participants punish their neighbors who adopt opposite strategies when their own payoffs are lower than their neighbors' average payoffs~\cite{32}. Wang {\it et~al.} proposed a tax-based pure reward and punishment approach and found that this approach has evolutionary advantages over a pure reward and punishment~\cite{33}. Such a brief summary cannot be complete because, as we noted, several other conditions were recognized and revealed which makes incentives effective or even more vulnerable~\cite{34,35,36,ohdaira_srep22,lv_amc22,flores_jtb21}.

It is undoubtedly that collective punishment has shown remarkable effectiveness in maintaining social cooperation and is widely used in human societies~\cite{37}. Generally speaking, citizens pay taxes to the government, and the government acts as the third party to punish the defectors and maintain social stability. However, in reality, not all members of the third party are fair, there are still instances of corruption and fraud in society. Recently, some researchers have focused on this problem. Verma and Sengupta developed deterministic and stochastic evolutionary game theory models of bribery and found that asymmetric punishment scenarios can reduce corruption under certain conditions~\cite{38}. Shi {\it et~al.} proposed a model of referee intervention involving corruption to explore the impact of corruption on the punishment mechanisms.  These results underlined that referee intervention always improves social efficiency, even in a completely corrupt system. Moreover, relatively developed institutions can resist corruption~\cite{39}. 

As the phenomenon of corruption exists, its impact on social dilemmas cannot be ignored. However, little attention has been paid to the benefits for the third-party adjudication. Inspired by this, we propose a multilayer network model approach that includes both a game layer formed by competing agents via a social dilemma and a monitoring layer occupied by referees who observe the former set. Due to the practical significance of multilayer networks, it often provides an appropriate topological framework for studying social dilemmas in evolutionary games \cite{wang_z_epl12,wang_z_epjb15,41,42}. In our model, nodes play the PGG with their neighbors in the game layer, which are monitored by the third-party nodes in the monitoring layer. We refer to these nodes as referees. However, not all of the referees are fair. Corrupt referees accept bribes from defectors without punishing them, while fair referees enforce penalties according to the rules. The amount of punishment is related to the payoffs that defectors receive in that iteration of the game. As a result, fair referees can receive their supervision fees, while corrupt referees focus on alternative sources. Their income originated from the bribes they have collected from defectors. It is an important aspect of our approach that not only the nodes in the game layer update their strategies, but also the behaviors of the nodes in the referee layer are dynamic and they also imitate their more successful partners. In this way, we establish the chance of a coevolutionary process where not only strategies but also judgment, meaning the status of supervisors evolve.
Through Monte Carlo simulations, we observe that the punishment ratio, the amount of bribe and the supervision fees all have a significant effect on the cooperative behavior under social dilemma. 

The rest of this paper is organized as follows. In Section~\ref{sec2}, we describe our model in detail, including the calculation of the payoffs of the nodes in the game layer and the individual incomes in referee layer. We also specify the microscopic update rules, which establish the coevolutionary link between the layers. In Section~\ref{sec3}, we outline the simulation methods, present our main findings and analyze their consequences. Finally, we conclude the research work of this paper and discuss future prospects in Section~\ref{sec4}.

\section{Model}\label{sec2}

In this section, we provide a detailed description of our model, including the network structure, the payoff calculations for different types of participants, and the strategy update rules.

\subsection{Game Model}

We construct a multilayer network structure consisting of a game layer and a referee layer, and suppose that each of which is a square lattice network with the same size. In the game layer, players play PGG with their nearest neighbors, hence forming a five-member group. The positions of players and referees in the network are matched, with each referee supervising the PGG centered on their corresponding players. We then briefly describe the PGG, where cooperators contribute some cost to the public pool, while defectors contribute nothing. The total investment in the public pool is multiplied by a synergy factor $r$ and then distributed equally to all participants, regardless of whether they are cooperators or defectors. Therefore, the PGG is a representative multiplayer social dilemma.

\subsection{Payoff Calculation}

In this paper, we denote the strategy of player (\textit{resp.} referee) $i$ at step $t$ as $x_i(t)$ (\textit{resp.} $y_i(t)$). For players, $x_i(t)\in\{0,1\}$, where $1$ and $0$ indicates a cooperator and a defector state respectively. For referees, $y_i(t) = 1$ if $i$ is a fair referee, otherwise $y_i(t)=0$. 

In our model, the population structure is assumed to be homogeneous. The payoff of player $i$ in the PGG organized by a neighbor $j$ at time step $t$ can be presented as
\begin{equation}\label{1}
P_{i,j}(t)=R_j-cx_i(t)-\alpha R_j (1-x_i(t))y_i(t)-\beta (1-x_i(t)) (1-y_i(t)), 
\end{equation}
where $R_j = \frac{r \sum\limits_{k \in \Omega_j} c x_k(t)}{\vert \Omega_j\vert}$, and $\Omega_j$ is the set consisting of the player $j$ and its neighbors. $c$ is the individual contribution of cooperators and $r$ is the synergy factor. $R_j$ is the payoff distributed equally to player $j$ and its neighbors in the game layer originated from the public pool. To prevent defection, a fair referee reduces the income of the defector with an $\alpha$ proportion. To avoid punishment defectors are willing to pay a bribe to a corrupt referee from their income.
Consequently, the total payoff of node $i$ at time step $t$ can be calculated as follows:
\begin{equation}\label{2}
    P_i(t) = \sum\limits_{j \in \Omega_i} P_{i,j}(t)
\end{equation}

The payoff for a referee $i$ at time step $t$ is
\begin{equation}\label{3}
\Pi_i(t)=my_i(t)+\beta\sum_{k\in\Omega_i}(1-x_k(t))(1-y_i(t))
\end{equation}
Hereby, the fair referees receive a certain supervision fee $m$ in each round. The corrupt referees turn to alternative sources and they collect bribes from defectors. Since the monitoring layer mirrors the topology of game layer, $\Omega_i$ represents the set of player $i$ and its neighbors in the game layer. All defectors within this set are required to pay a bribe of $\beta$ to the corrupt referee, which means that the more defectors, the larger amount of bribe can be collected by a corrupt referee.

\subsection{Strategy Evolution}

Individuals update their strategies via an imitation process after obtaining their accumulated payoffs. In particular, we use the so-called imitation with pairwise comparison where the imitation probability is calculated from Fermi function. Accordingly, individuals are more likely to adopt the strategy of those who have higher payoffs than themselves, but ``irrational'' decision is also possible with low probability. It is an important feature of our model that individuals in the referee layer can also update their strategies over time for a higher payoff and fitness. In this process, some fair referees will give up their principles and take bribes to become corrupted for higher payoffs, while others do not change their nature for the sake of immediate benefits and still insist to be the fair referees. Unlike the two strategies of cooperation and defection in the game layer, we use ``fair'' and ``corrupt'' to describe the behaviors of referees in the referee layer. Therefore, in both the game and the referee layers, the vertex $i$ randomly chooses a neighbor $j$ at each time step, and compares their payoffs in this round. The probability that $i$ adopts $j$'s strategy in the next time step is given by
\begin{equation}\label{4}
    W(S_i \leftarrow S_j)=\frac{1}{1+e^{(P_i - \ P_j) / \kappa}},
\end{equation}
where $S_i$ and $P_i$ represent the strategy and payoff of node $i$ respectively. Evidently, $P_i$ (\textit{resp.} $P_j$) is replaced by $\Pi_i$ (\textit{resp.} $\Pi_j$) in the referee layer.
The strategy of $S_i$ is cooperation or defection if it is the game layer, and it is fair or corrupt for the referee layer. $\kappa$ represents the noise level that characterizes the chance of irrational choices during the microscopic dynamics. Specifically, the lower
the payoff of $i$ compared with $j$, the more likely that $i$ adopts $j$'s strategy, and vice versa. Importantly, members on the game layer and on the referee layer update their status independently at each time step, which means that both player $i$ and referee $i$ will each perform their own strategy update respectively.
\begin{figure}
    \centering
    \includegraphics[width=0.95\linewidth]{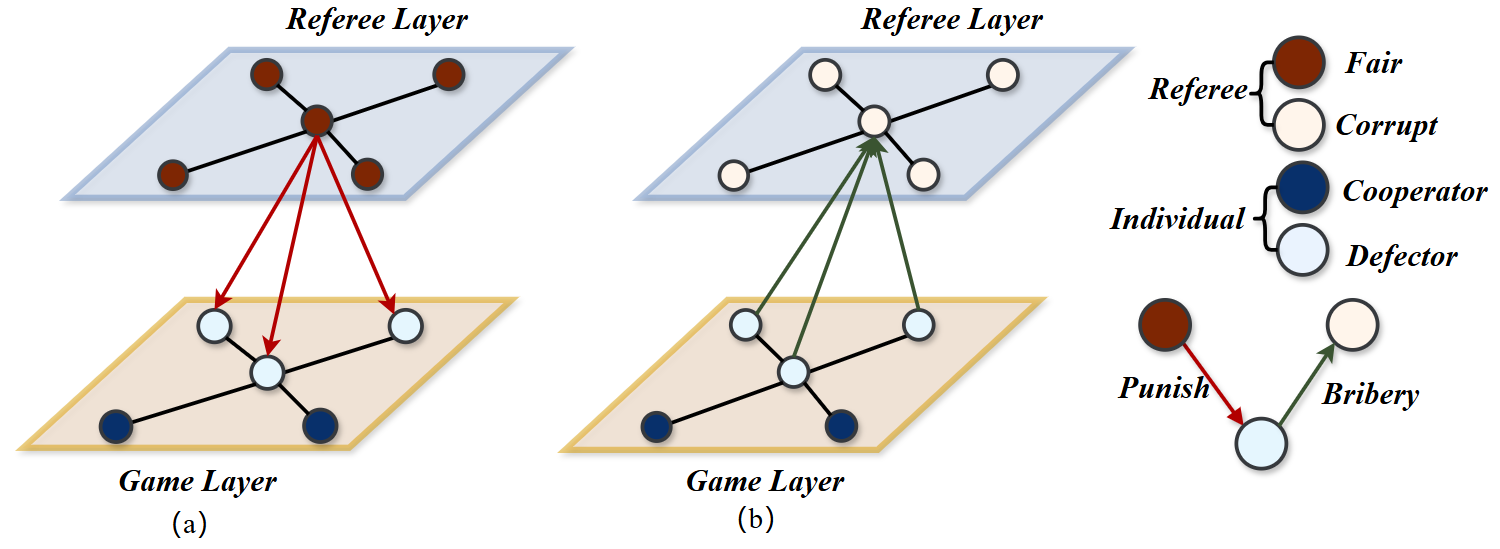}
    \caption{\textbf{Interdependent approach to model coevolution between strategies and judgments.} The lower layer is the stage of social dilemma where cooperator and defector players interact via spatial public goods game. The upper layer hosts supervisors or referees who monitor the strategy evolution and mediate it indirectly. Panel~(a) illustrates the case when a fair referee punishes defectors in the corresponding group. Panel~(b) shows defectors' response who pay a bribe to a corrupt referee hence avoid being punished.}
    \label{fig: example}
\end{figure}

To better illustrate our model, we briefly summarize the key elements in Fig.~\ref{fig: example}. As noted we use a multilayer network approach and divide it into a game layer and a referee layer, with nodes in the two layers are linked. Nodes in the game layer play the PGG, while nodes in the referee layer will supervise these games. Accordingly, fair referees punish defectors in the corresponding group and receive certain supervision fees, while corrupt referees accept bribes from related defectors and help them to avoid punishment. Consequently, corrupt referees cannot receive supervision fees, and their payoffs come from the amounts of bribes they collect. For strategy updating, we consider that both players and referees will update their strategies via an imitation process based on pairwise comparison of payoff values. 

\section{Simulation results and analyses}\label{sec3}

In this section, we conduct Monte Carlo simulations to validate the model we proposed. We study the effects of supervision fees, penalty ratios and bribe costs on the frequency of cooperation. We also focus on the referee layer to explore the potential connection between the pattern formations due to co-evolutionary process. Finally, we reveal the resulting payoff distribution both for referees and players.

\subsection{Methods}

In the initial stage of each simulation, we use a square lattice network with the periodic boundary conditions for both layers where the linear system size is $L=50$. This means that there are a total of 2500 nodes in both game and referee layer. Initially, each node is randomly assigned as a cooperator (fair referee) or a defector (corrupt referee), with an equal probability. In each Monte Carlo step, on average all participants randomly select a neighbor to compare their payoffs and decide whether to learn the strategy of the neighbor with a probability given in Eq.~\ref{4}. According to our observations, the evolution process always reaches a stable state after 2500 time steps. Therefore, we calculate the $f_C$ cooperation density and the $f_F$ frequency of fairness by averaging the last 500
steps of 3000 total steps. To achieve the requested accuracy of our simulations, each result obtained at a particular set of parameters is averaged over 10 independent simulations. For simplicity, we set $c=1$ and $\kappa=0.1$. The latter noise value ensures that it is more likely to adopt a strategy with higher income, but the reverse process is also possible, albeit with low probability.

\subsection{Effect of penalty ratio $\alpha$ on cooperation and fairness}

\begin{figure*}[t]
    \centering
    \begin{subfigure}{0.42\linewidth}
        \centering
        \includegraphics[width=\linewidth]{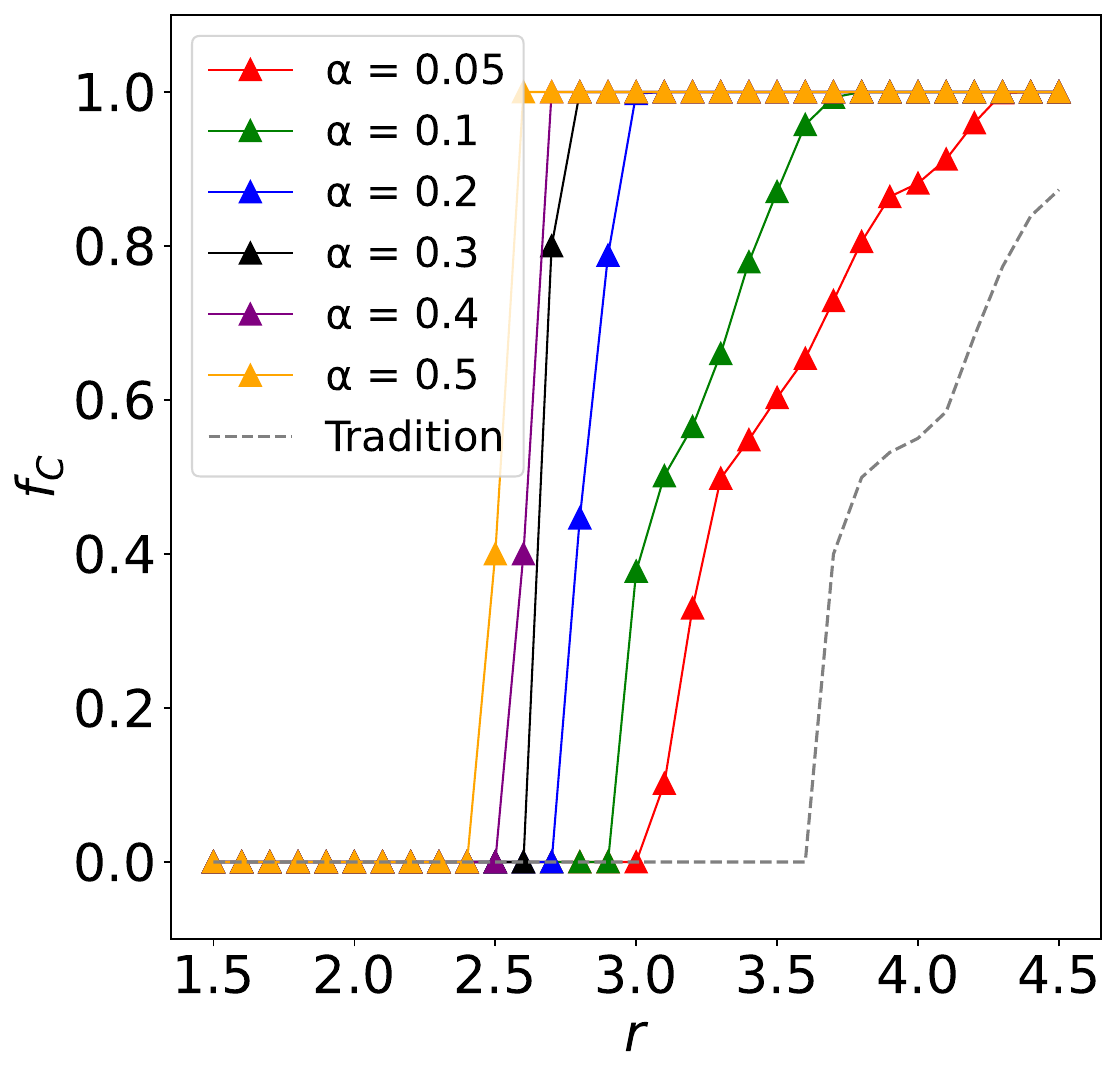}
        \caption{}
        \label{fig2:a}
    \end{subfigure}
    \hfill
    \begin{subfigure}{0.42\linewidth}
        \centering
        \includegraphics[width=\linewidth]{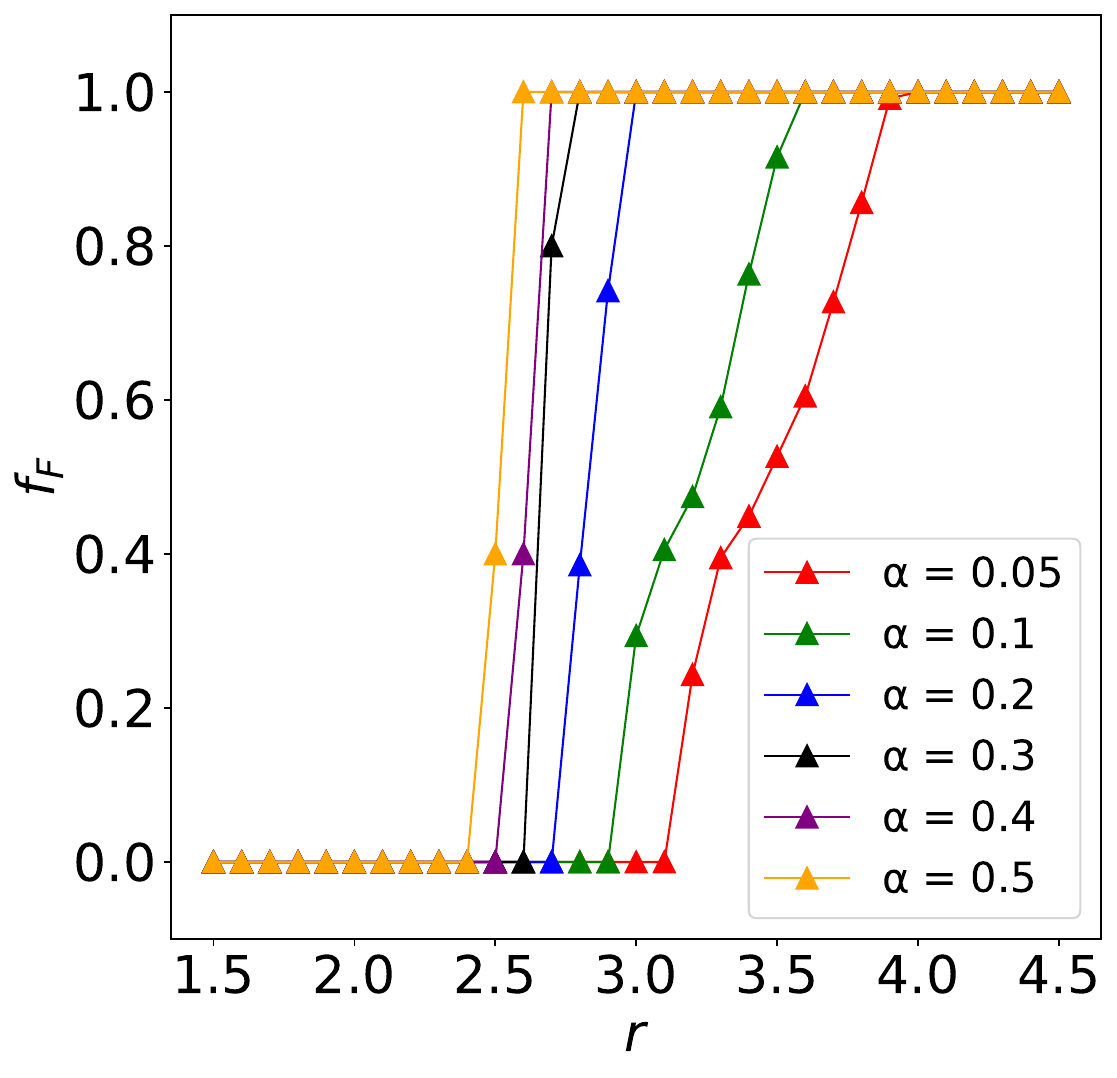}
        \caption{}
        \label{fig2:b}
    \end{subfigure}
    \caption{\textbf{Cooperation level and the portion of fair supervisors in dependence of synergy factor at different fine values.} Panel~(a) shows the fraction of cooperators in the game layer as we increase the synergy factor $r$. The values of fine for defectors are indicated in the legend. Panel~(b) depicts the fraction of fair referees in the monitoring layer. Other parameters, the supervision fee $m=0.5$ and the bribery cost $\beta=0.2$ are fixed. The stationary values are calculated from the last 500 steps of 3000 total steps and averaged over 10 independent simulations. The gray dashed line in Panel~(a) indicates the baseline without referees. } 
    \label{fig2}
\end{figure*}

The cooperation level usually attracts interest and attention, and the cooperation density $f_C$, a common indicator of global cooperation level, represents the proportion of cooperation. In this paper, we also focus on the referee layer, using $f_F$ to measure the proportion of fair referees in the referee layer. 

To focus on the impact of the punishment ratio $\alpha$ on $f_C$ and $f_F$, we plot the curves of cooperation density $f_C$ and the proportion of fair referees $f_F$ against the coefficient of synergy $r$ under different $\alpha$ values in Fig.~\ref{fig2}. From this figure, we can find that no matter whether it is the game layer or the referee layer, all the curves first maintain the pure defection or corruption state for small $r$ values which represent harsh conditions for cooperation. After both $f_C$ and $f_F$ rise rapidly with the increase of $r$, and finally reach the state of pure cooperation and fairness, respectively. Evidently, the highest $\alpha=0.5$ fine portion provides the best condition for the pure cooperation and pure fairness states. On the contrary, $\alpha=0.05$ provides the narrowest parameter space for the pure cooperation and pure fairness states, but it provides the widest parameter range for a mixed strategy state. For $\alpha=0.05$, the pure defection and pure corruption states last until $r=3.0$ and $r=3.1$ respectively. It is worth noting, however, that this threshold value is still significantly lower than $r_c=3.744$ previously found in the traditional PGG model on the same topology~\cite{szolnoki_pre09c}. For $\alpha=0.1$ and $\alpha=0.2$, the thresholds for the existence of cooperators in the game layer are $r=2.9$ and $r=2.7$, and the same thresholds apply in the referee layer. For $\alpha=0.3$ and $\alpha=0.4$, the thresholds for the existence of cooperators and fair referees are $r=2.5$ and $r=2.6$. However, as $\alpha$ increases, there is a rapid phase transition whether it is the from the pure defection to the pure cooperation state in the game layer, or the pure corruption to the pure fairness state in the referee layer. When $\alpha=0.5$, the thresholds for the pure defection state and pure cooperation state in the game layer are $r=2.4$ and $r=2.6$ respectively, and the thresholds for the pure corruption state and pure fairness state in the referee layer are the same as above. The reason is that when $\alpha$ is very large, players who choose to defect need to pay a serious penalty, which makes it easier for the cooperators to prevail in the game, and the penalty cost $\alpha$ has a greater impact on the evolutionary process at this time, leading to the rapid phase transitions in the curves. This is also consistent with reality, as the government can effectively reduce the occurrence of such behaviors by appropriately increasing the punishment on illegal activities.

\subsection{Effect of bribery cost $\beta$ on $f_C$ and $f_F$}

\begin{figure*}[t]
    \centering
    \begin{subfigure}{0.42\linewidth}
        \centering
        \includegraphics[width=\linewidth]{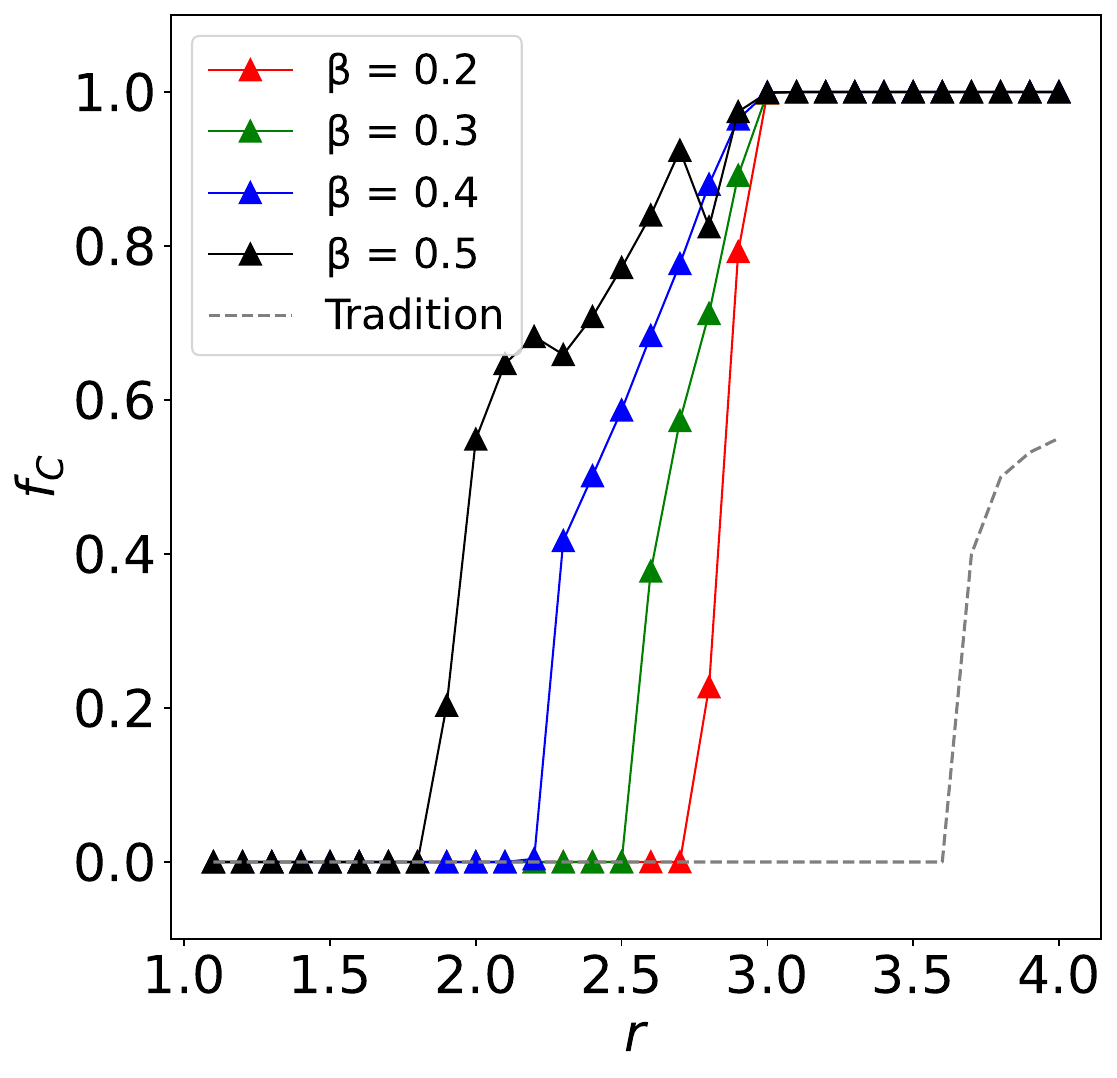}
        \caption{}
        \label{fig3:a}
    \end{subfigure}
    \hfill
    \begin{subfigure}{0.42\linewidth}
        \centering
        \includegraphics[width=\linewidth]{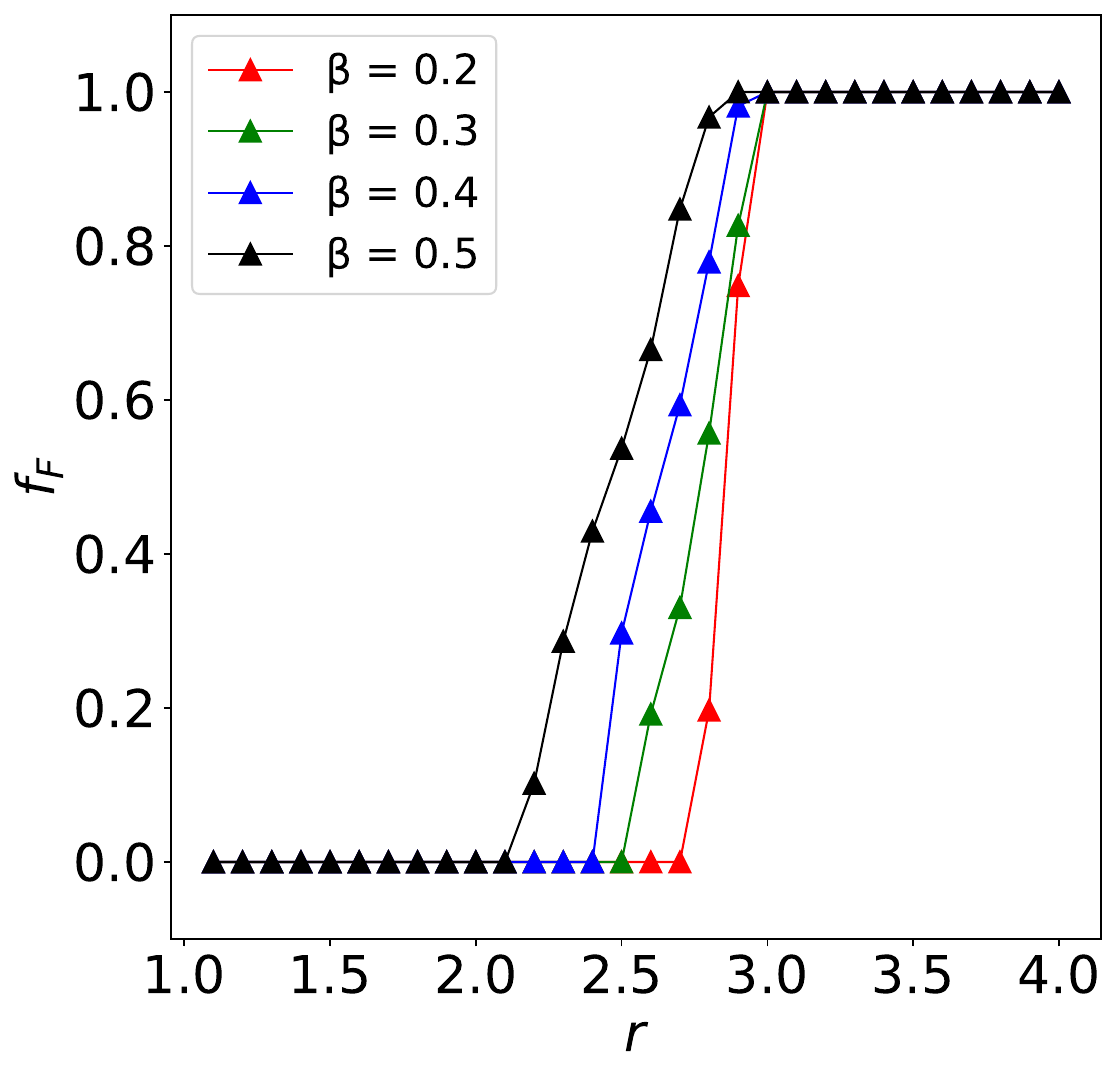}
        \caption{}
        \label{fig3:b}
    \end{subfigure}
    \caption{\textbf{Cooperation level and the portion of fair supervisors in dependence of synergy factor at different bribe values.} Panel~(a) shows the fraction of cooperators in the game layer, while panel~(b) depicts the fraction of fair referees in the monitoring layer. The values of bribe paid by defectors to avoid punishment are marked in the legend. The remaining parameters, the supervision fee $m=0.5$ and the punishment level $\alpha=0.2$ are fixed. The stationary values are calculated from the last 500 steps of 3000 total steps and averaged over 10 independent simulations. The gray dashed line in Panel~(a) indicates the baseline without referees. } 
    \label{fig3}
\end{figure*}

Next, in order to focus on the effect of bribery cost $\beta$ on agents’ behaviors, we plot the curves of cooperation density $f_C$ and the proportion of fair referees $f_F$ in dependence of synergy factor $r$ in Fig.~\ref{3}. We find that both $f_C$ and $f_F$ exhibit roughly the same trend when other parameters are fixed. As previously, all curves terminate into the pure defection or pure corruption state when synergy factor $r$ is low, indicating that this parameter plays the key role and its principal impact cannot be overridden with appropriate fines or bribes. The opposite is also true because full cooperation and pure fairness state can also be reached if the synergy factor is high.
Additionally, the larger the value of $\beta$, the earlier the appearance of cooperators and fair referees, and the earlier the transition to the pure cooperation and pure fairness states, in both the game layer and the referee layer. It can be observed that the state of pure defection and pure corruption last until $r=2.7$ which is the longest when $\beta=0.2$. In other words, cooperators and fair referees only appear after that. Interestingly, however, even such a small bribe is not enough for defectors to expand their kingdom, because this threshold value is significantly lower than the mentioned $r_c=3.745$ for the bribe-free, traditional PGG model. For $\beta=0.3$ and $\beta=0.4$, the thresholds for the emergence of cooperators are $r=2.5$ and $r=2.2$, and the thresholds for the appearance of fair referees are $r=2.5$ and $r=2.4$, both lower than the case of $\beta=0.2$. When $\beta=0.5$, there exists the lowest threshold for the pure defection and corruption, providing the largest parameter space for the emergence of cooperators and fair referees. In addition, their defection and corruption annihilation thresholds are similar for different $\beta$, which implies that both $f_C$ and $f_F$ reach its saturation level when $r$ increases to 3.0. By comparing Fig.~\ref{fig3:a} and Fig.~\ref{fig3:b}, we also find that the thresholds for the emergence of cooperator are always lower than those for the fair referee, and the larger the $\beta$, the greater the gap between the two thresholds. This is because cooperators will gradually take the dominant position as $r$ increases. However, it is inevitable that defectors still dominate the game layer for small $r$ and corrupt referees in the monitoring layer can benefit from it. They earn larger incomes than fair referees who eventually adopt the more profitable attitude. This will result in a zero $f_F$ value in the referee layer. The larger the $\beta$, the more bribes are collected by corrupt referees. It means that more cooperators need to emerge in the game layer in order for corrupt referees to collect lower bribes than the supervision fee of fair referees. This will also lead to a wider gap between the threshold for the appearance of fair referees and cooperators.

\subsection{The cooperation and fairness landscapes on punishment-bribe parameter plane}

\begin{figure*}[h!]
    \centering
    \begin{subfigure}[b]{0.4\textwidth}
        \centering
        \includegraphics[width=\textwidth]{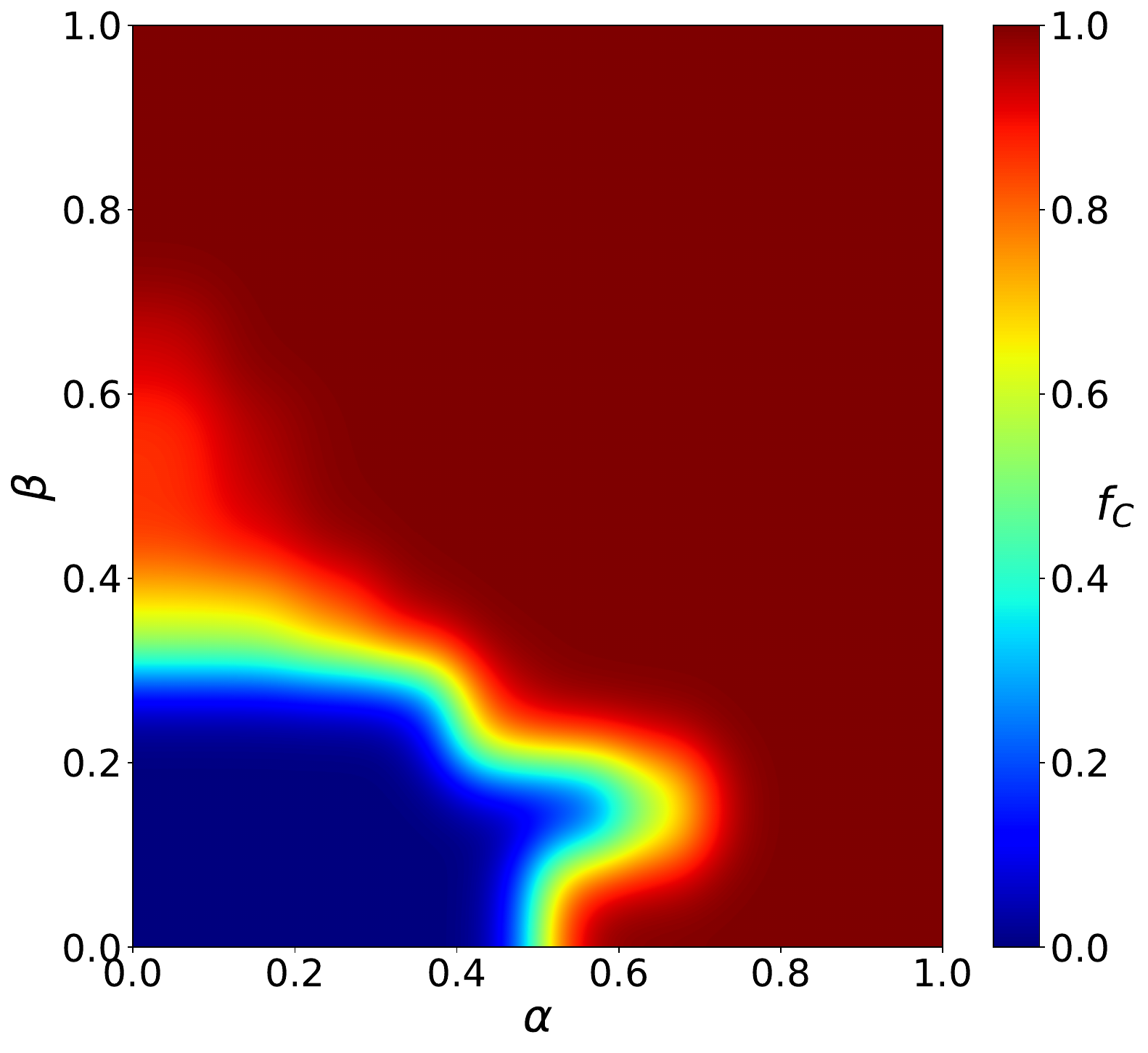}
        \caption{}
        \label{fig4:a}
    \end{subfigure}
    \hspace{-3mm}
    \begin{subfigure}[b]{0.4\textwidth}
        \centering
        \includegraphics[width=\textwidth]{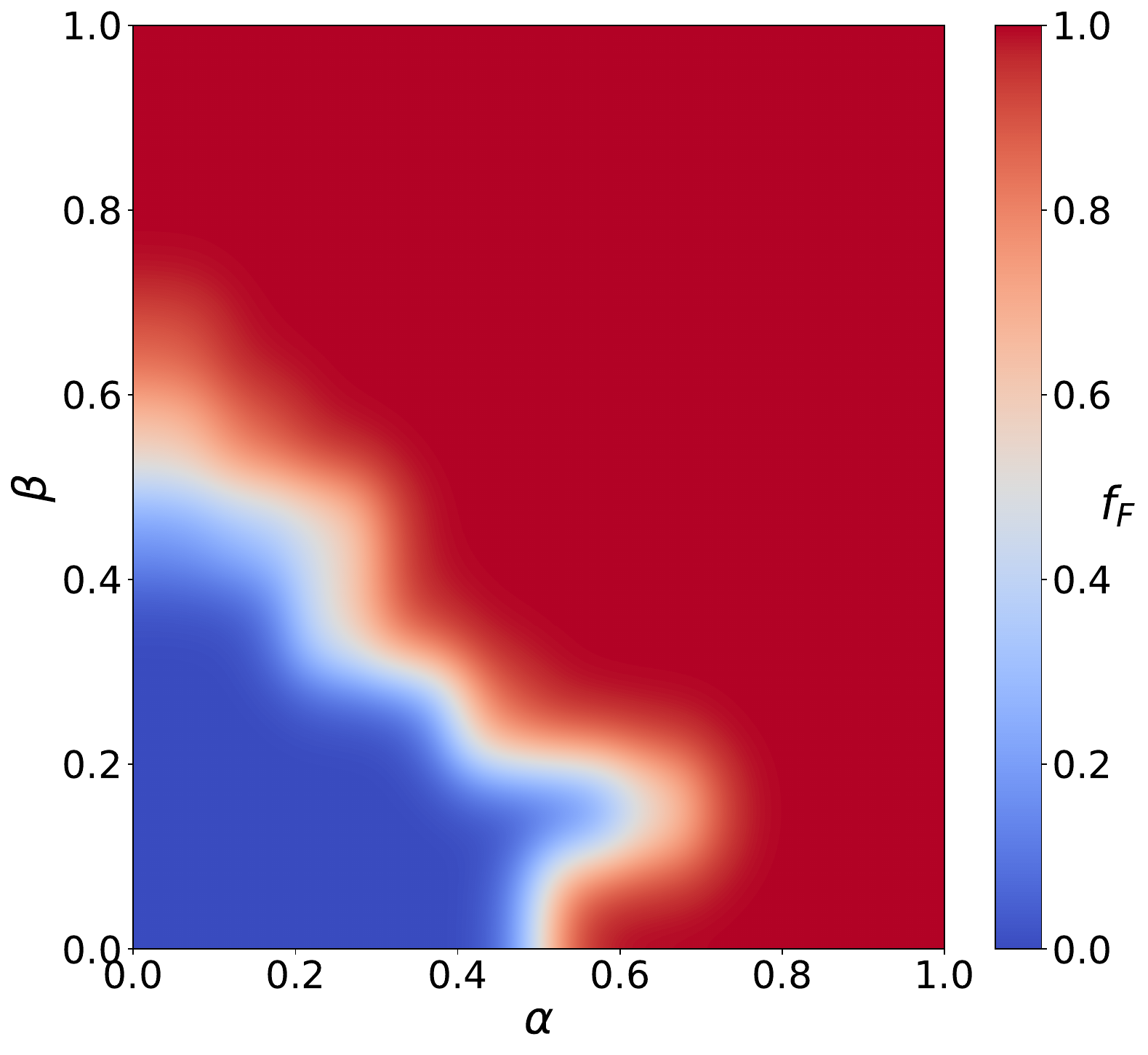}
        \caption{}
        \label{fig4:b}
    \end{subfigure}

    \begin{subfigure}[b]{0.4\textwidth}
        \centering
        \includegraphics[width=\textwidth]{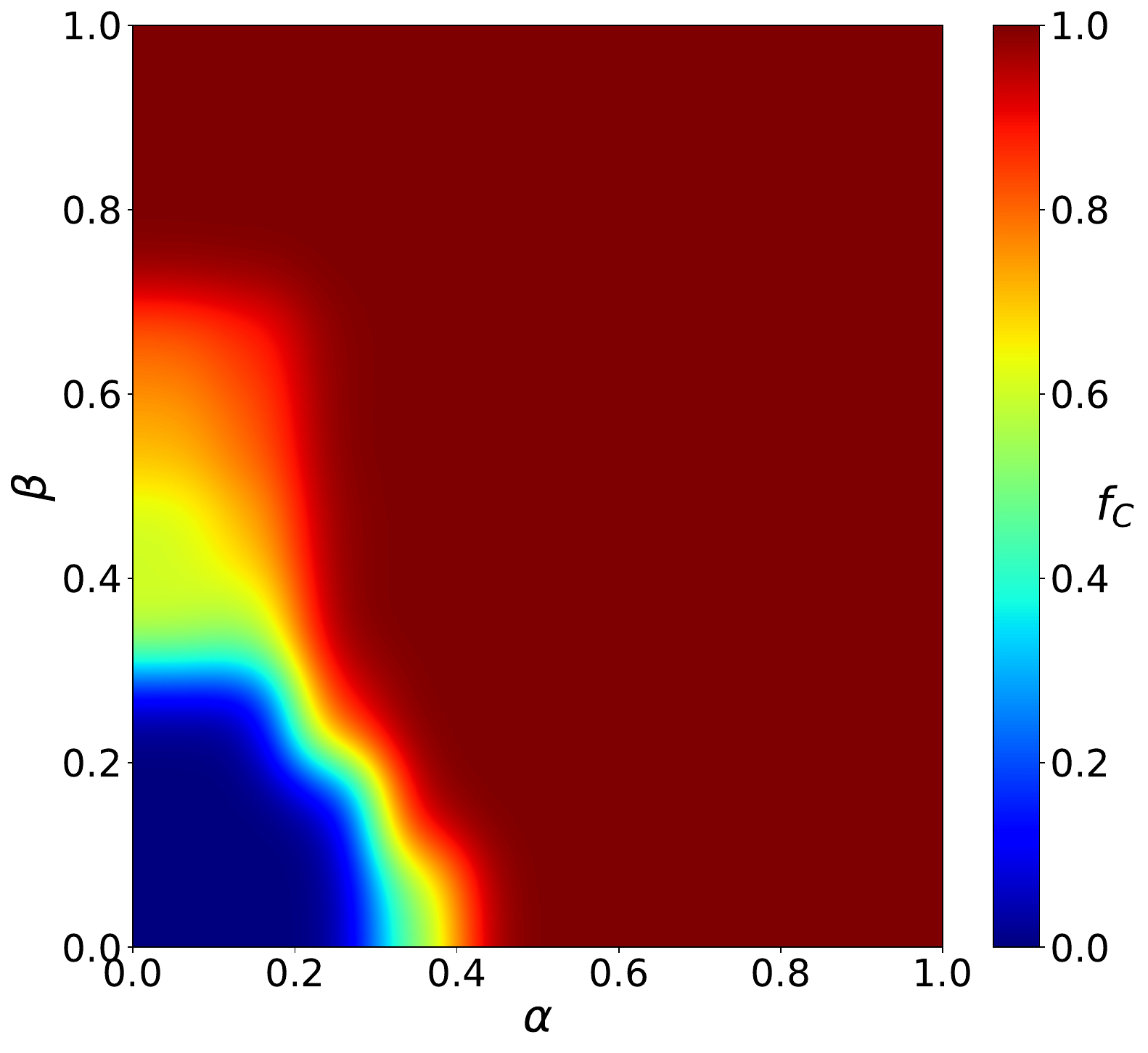}
        \caption{}
        \label{fig4:c}
    \end{subfigure}
    \hspace{-3mm}
    \begin{subfigure}[b]{0.4\textwidth}
        \centering
        \includegraphics[width=\textwidth]{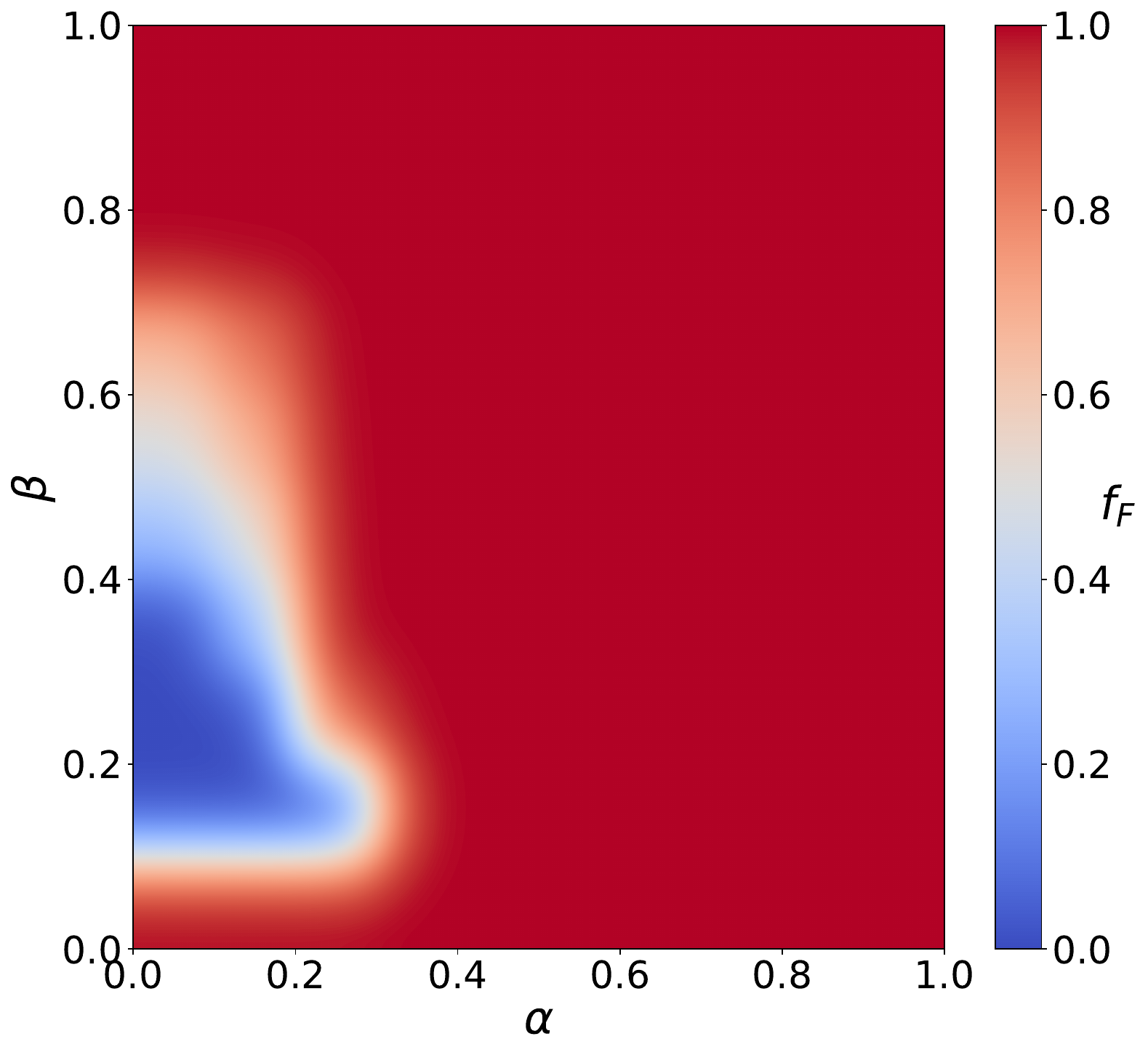}
        \caption{}
        \label{fig4:d}
    \end{subfigure}
    \vspace{-3mm}
    
    \caption{\textbf{The consequence of heterogeneous supervisors on fine-bribery parameter plane.} The left panels show the color-coded portion of cooperators in the game layer on $\alpha-\beta$ parameter plane. Right panels show the color coded density of fair referees in the monitoring layer at the same fine-bribery parameter pairs. Top row shows the case obtained at $m=0.3$ supervision fee, while bottom row indicates when $m=0.7$. The synergy factor was fixed $r=2.5$ for all cases. The cooperation frequency for each data point is averaged from the last 500 steps of 3000 total steps and averaged over 10 independent simulations. }
    \label{fig4}
\end{figure*}

To explore the combined effects of the punishment ratio $\alpha$ and the bribery cost $\beta$ on the evolution more generally we summarize the key quantities on the mentioned parameter plane shown in Fig.~\ref{fig4}. The left-hand panels show the general cooperation level for two representative values of supervision fee. The right-hand panels show the related concentration of fairness in the monitoring layer. The values are color-coded as indicated in the legend attached to each panel. As a general observation, both $f_C$ and $f_F$ grow when we increase the punishment ratio or the bribery cost. For each specified $\alpha$, as $\beta$ increases, both $f_C$ and $f_F$ rise, which is consistent with our previous analysis, and similar agreement is found when $\beta$ is fixed while $\alpha$ varies. Meanwhile, the cooperative density and the level of fairness show similar trends on the heatmap. As shown in Fig.~\ref{fig4}, the bottom-left regions of the four heatmaps are colored in blue, indicating that neither the cooperation density nor fairness cannot be maintained in the absence of incentives at such a low value of synergy factor. In contrast, the top-right regions always exhibit the opposite. We find that the game layer gets rid of the pure defection state when $\beta > 0.3$ in Fig.~\ref{fig4:a}, while the referee layer is still in a pure corruption state at these parameter pairs, as shown in Fig.~\ref{fig4:b}. A similar trend can be observed in Fig.~\ref{fig4:c} and Fig.~\ref{fig4:d} when fair referees are awarded by a significantly high supervision fee. This is because when $\alpha$ is small, as the bribery cost $\beta$ increases, the defectors in the game layer need to pay more bribes to the corrupt referees in each round to ensure their immunity from punishment. Here, the payoffs of some cooperators will be higher than those of the defectors, allowing cooperators to emerge. In the referee layer, the larger $\beta$ involves higher bribe for corrupt referees, while fair referees should make do with a modest supervision fee. Consequently, the referee layer is still dominated by corrupt referees because their payoffs are higher than those of fair referees. Meanwhile, we observe that when both $\alpha$ and $\beta$ are very small, the $f_F$ portion of fair referees is 0 in Fig.~\ref{fig4:b}, while Fig.~\ref{fig4:d} shows the opposite situation with the $f_F$ reaching 1. This is mainly due to the higher supervision fee $m$ in the latter case. Note that the supervision fee is significantly lower ($m=0.3$) in Fig.~\ref{fig4:b}. When both $\alpha$ and $\beta$ are very small $(\alpha = 0.1, \beta = 0.1)$, the game layer is full of defectors because of the lack of punishment. Here, corrupt referees can collect bribes of 0.5, which means a significantly higher income to them than the payoff of fair referees. Consequently, the applied imitation dynamics will result in the prevalence of corruption. In contrast, in Fig.~\ref{fig4:d}, the supervision fee is pretty large. Even if there are all defectors in the game layer, the payoffs for fair referees remain higher than those for corrupt referees, leading to the referee layer being composed entirely of fair referees. This is also consistent with the real world, where appropriately increasing the income of government officials can effectively reduce the occurrence of corruption.

\subsection{Coevolutionary pattern formation of cooperation and fairness}

\begin{figure*}[h!]
    \centering
    \begin{subfigure}[b]{0.24\textwidth}
        \centering
        \includegraphics[width=\textwidth]{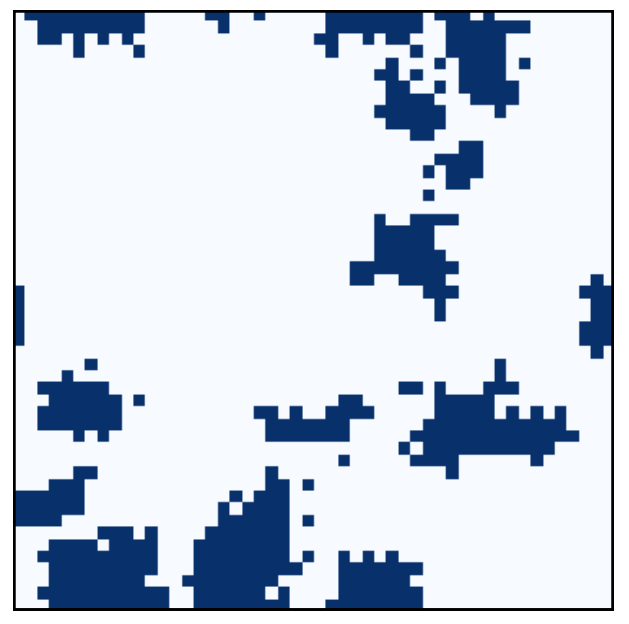}
        \caption{}
        \label{fig5:a}
    \end{subfigure}
    \hfill
    \begin{subfigure}[b]{0.24\textwidth}
        \centering
        \includegraphics[width=\textwidth]{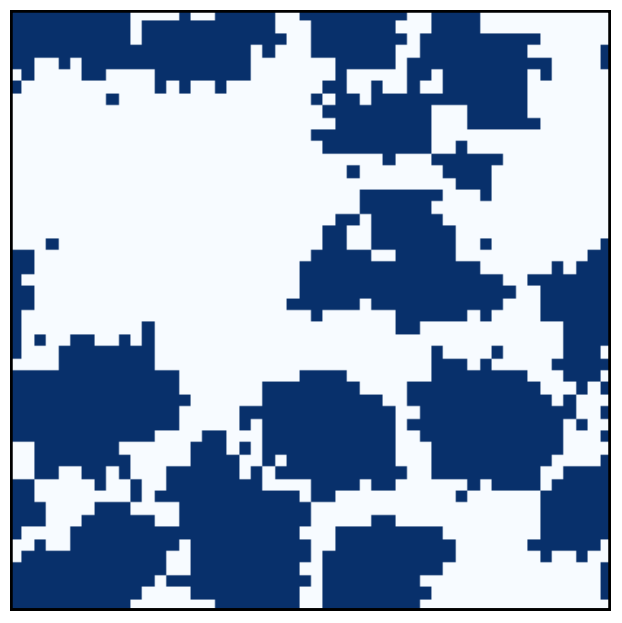}
        \caption{}
        \label{fig5:b}
    \end{subfigure}
    \hfill
    \begin{subfigure}[b]{0.24\textwidth}
        \centering
        \includegraphics[width=\textwidth]{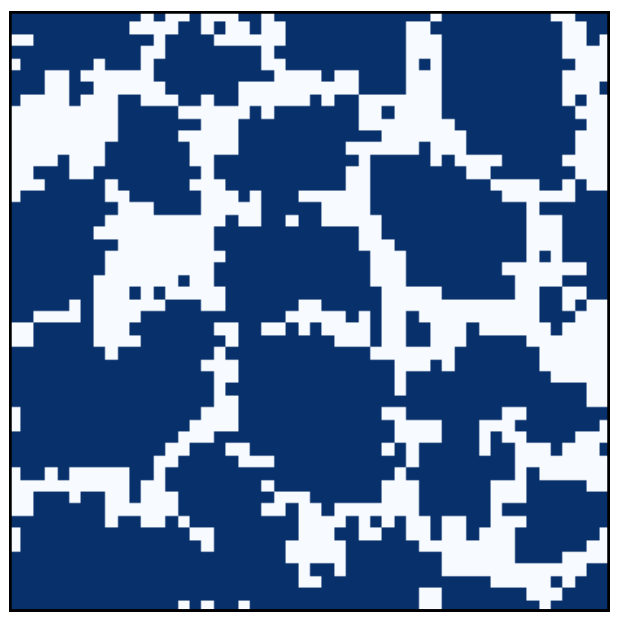}
        \caption{}
        \label{fig5:c}
    \end{subfigure}
    \hfill
    \begin{subfigure}[b]{0.24\textwidth}
        \centering
        \includegraphics[width=\textwidth]{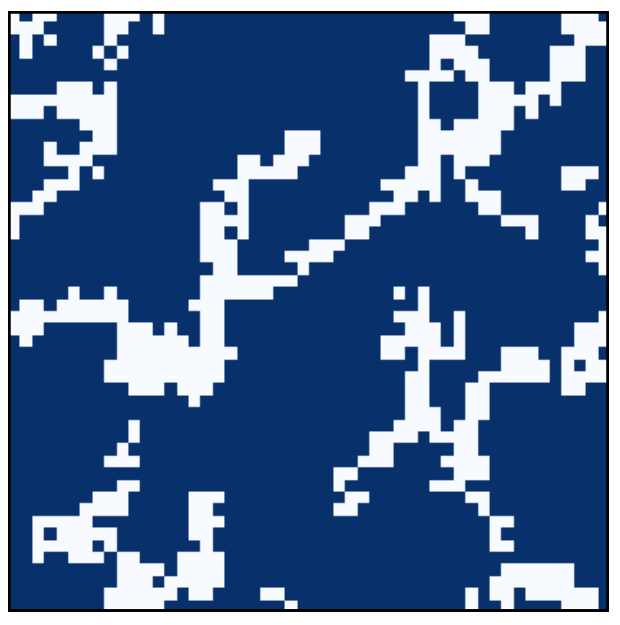}
        \caption{}
        \label{fig5:d}
    \end{subfigure}

    \vspace{\baselineskip}

    \begin{subfigure}[b]{0.24\textwidth}
        \centering
        \includegraphics[width=\textwidth]{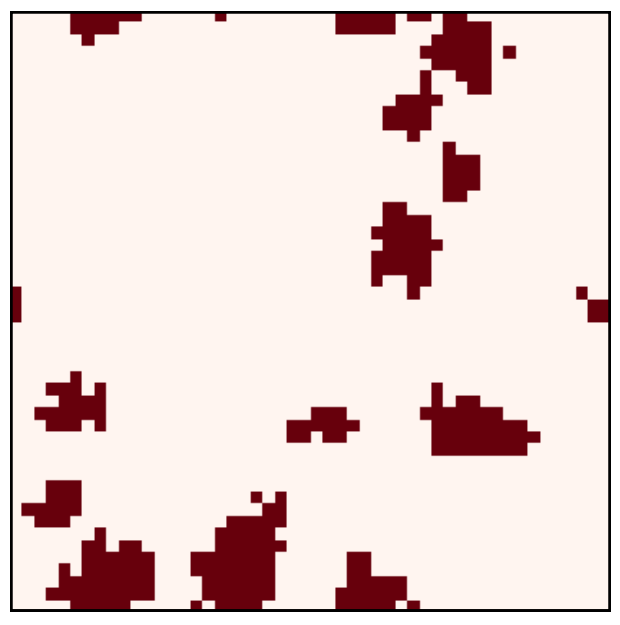}
        \caption{}
        \label{fig5:e}
    \end{subfigure}
    \hfill
    \begin{subfigure}[b]{0.24\textwidth}
        \centering
        \includegraphics[width=\textwidth]{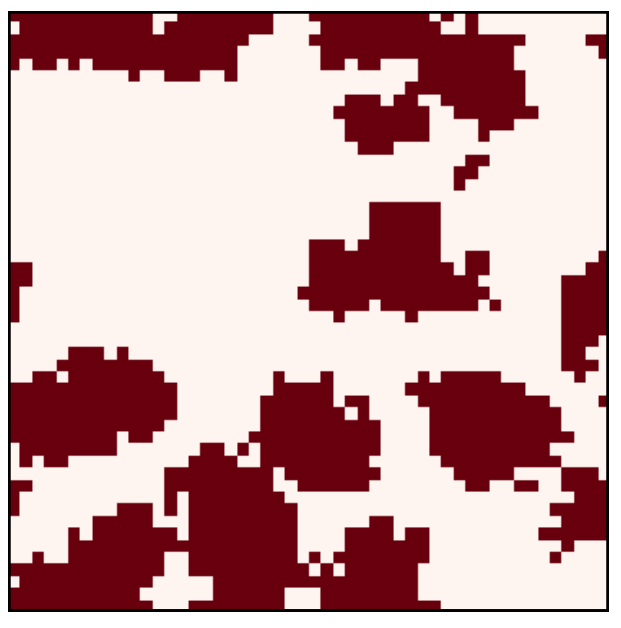}
        \caption{}
        \label{fig5:f}
    \end{subfigure}
    \hfill
    \begin{subfigure}[b]{0.24\textwidth}
        \centering
        \includegraphics[width=\textwidth]{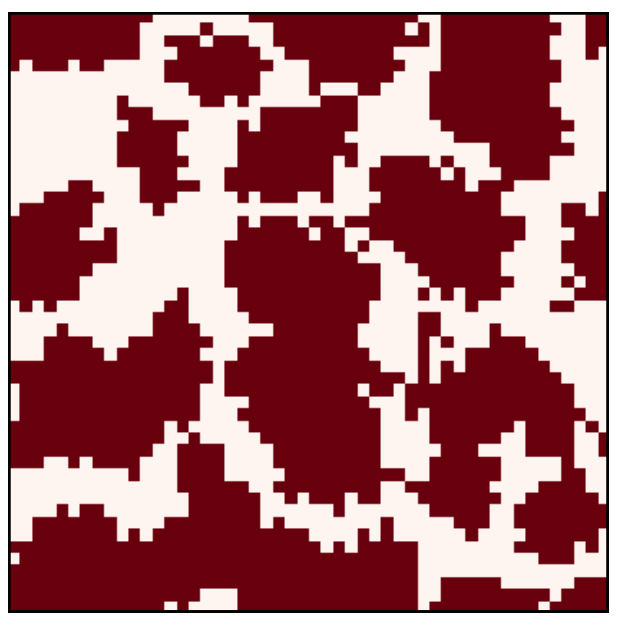}
        \caption{}
        \label{fig5:g}
    \end{subfigure}
    \hfill
    \begin{subfigure}[b]{0.24\textwidth}
        \centering
        \includegraphics[width=\textwidth]{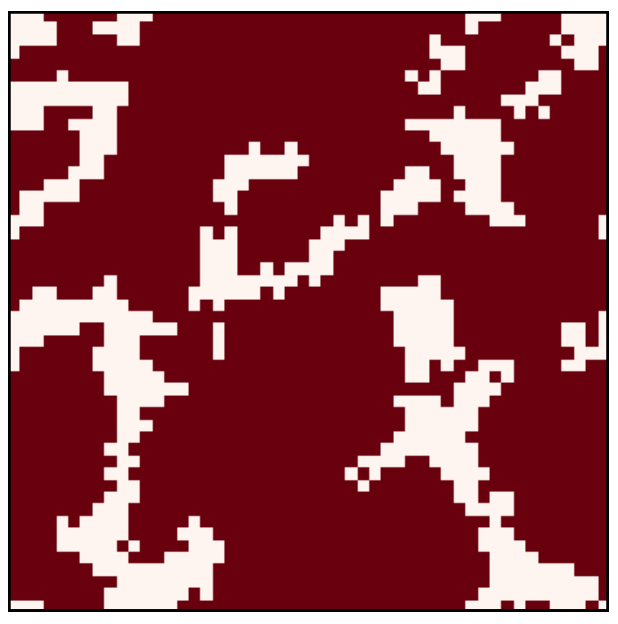}
        \caption{}
        \label{fig5:h}
    \end{subfigure}
    \caption{\textbf{Coevolution of cooperation and fairness.} Starting from a random initial state panels~(a)-(d) show the actual distribution of strategies in the game layer obtained at $t\in\{20,50,100,1000\}$ respectively. Here blue (\textit{resp.} white) cells represent cooperator (\textit{resp.} defector) agents. In the bottom row panels~(e)-(h) denote the distribution of competing referees in the monitoring layer obtained at the same steps specified above. Here red (\textit{resp.} linen) color represents fair (\textit{resp.} corrupt) referees. Other parameters are $r=3.4$, $m=0.5$, $\alpha = 0.1$, and $\beta = 0.2$. As the panels show, cooperation and fairness evolve hand in hand by supporting each other.}
    \label{fig5}
\end{figure*}

To understand the interdependence of cooperation and fairness more deeply we study the pattern formation on both layers simultaneously. In Fig.~\ref{fig5} we present how the distribution of agents evolves in time both in the game layer and in the monitoring layer. More precisely, starting from a random initial state we recorded the actual distributions obtained after 20, 50, 100, and 1000 MC steps. At the very beginning, shown in panel~(a) and in panel~(e), an undesired process happens in both layers. More precisely, defectors in the game layer and corrupt referees in the monitoring layer occupy the majority of space because random initial state offers an ideal condition for selfishness. More precisely, when cooperators and defectors are evenly distributed in the game layer, defectors can obtain higher payoffs from the public pool than the fines or bribes they need to pay. Therefore, cooperators learn the strategies of defectors, leading to an increase in defectors in the game layer. In the referee layer, corrupt referees can collect more bribes from defectors, resulting in the proportion of corrupt referees to rise.
Later, surviving clusters of cooperators and fair referees start growing because defectors can collect just a modest income from the public pool, which makes them difficult to pay fines or bribes. The advantage of the cooperators gradually becomes apparent, which will lead to a decrease in defection. Consequently, the bribes collected by corrupt referees in the monitoring layer also diminish, making the income of fair referees more attractive, ultimately allowing them to dominate the network. This process reminds us of the well-known dynamics of network reciprocity first reported in Refs.~\cite{perc_pre08b,szolnoki_epjb09}. 

What is more spectacular is the mentioned domains in separate layers grow in a correlated way. More precisely, if cooperators invade a certain area in the game layer then the corresponding area will be occupied by fair referees in the monitoring layer. In this way the mentioned actors can support each other across the layers and cooperation and fairness can evolve hand in hand. We note that the reported effect is a more general consequence of the applied topology and it is also called as interdependent network reciprocity in other evolutionary game models~\cite{szolnoki_njp13,wang_z_srep13}.

\subsection{Payoffs of game and referee layers}

\begin{figure*}[h!]
    \centering
    \begin{subfigure}[b]{0.24\textwidth}
        \centering
        \includegraphics[width=\textwidth]{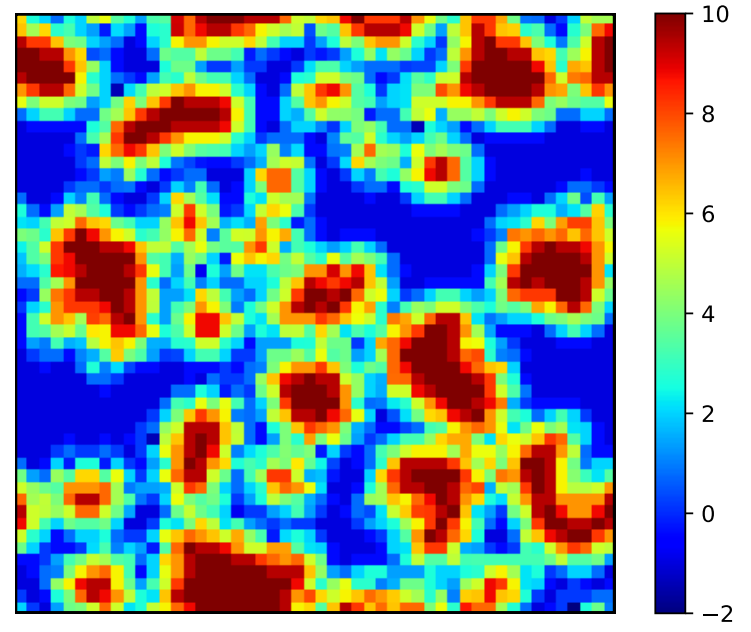}
        \caption{}
        \label{fig6:a}
    \end{subfigure}
    \hfill
    \begin{subfigure}[b]{0.24\textwidth}
        \centering
        \includegraphics[width=\textwidth]{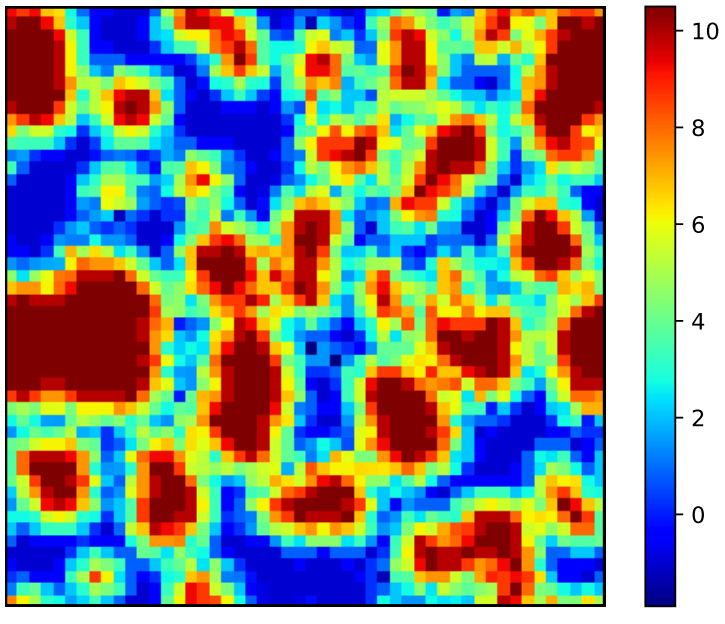}
        \caption{}
        \label{fig6:b}
    \end{subfigure}
    \hfill
    \begin{subfigure}[b]{0.24\textwidth}
        \centering
        \includegraphics[width=\textwidth]{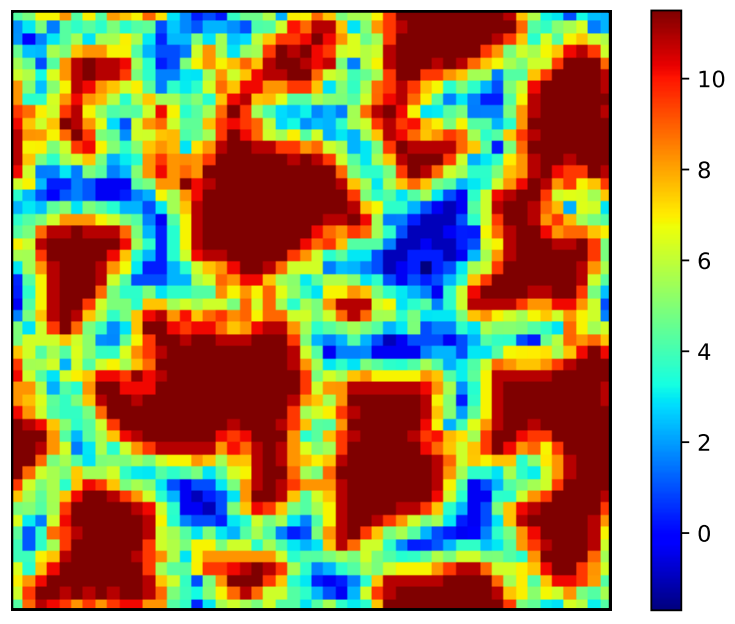}
        \caption{}
        \label{fig6:c}
    \end{subfigure}
    \hfill
    \begin{subfigure}[b]{0.24\textwidth}
        \centering
        \includegraphics[width=\textwidth]{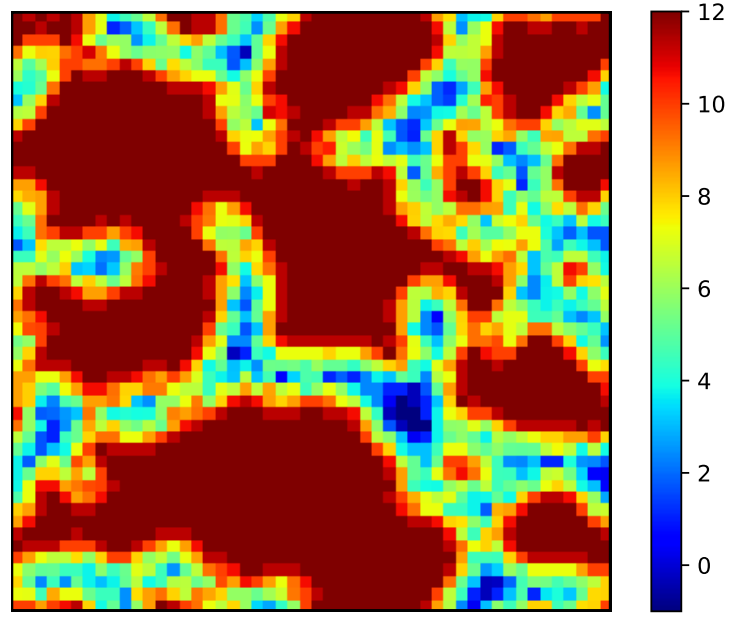}
        \caption{}
        \label{fig6:d}
    \end{subfigure}

    \vspace{\baselineskip}

    \begin{subfigure}[b]{0.24\textwidth}
        \centering
        \includegraphics[width=\textwidth]{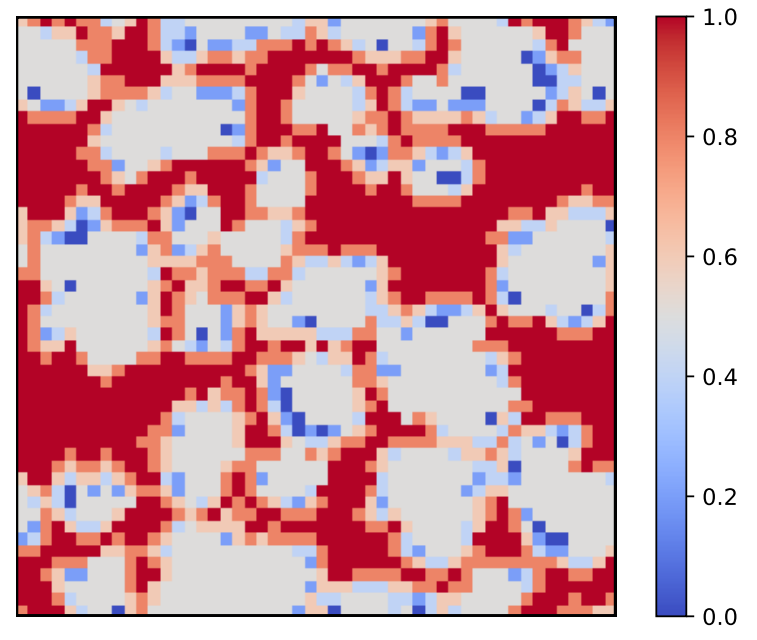}
        \caption{}
        \label{fig6:e}
    \end{subfigure}
    \hfill
    \begin{subfigure}[b]{0.24\textwidth}
        \centering
        \includegraphics[width=\textwidth]{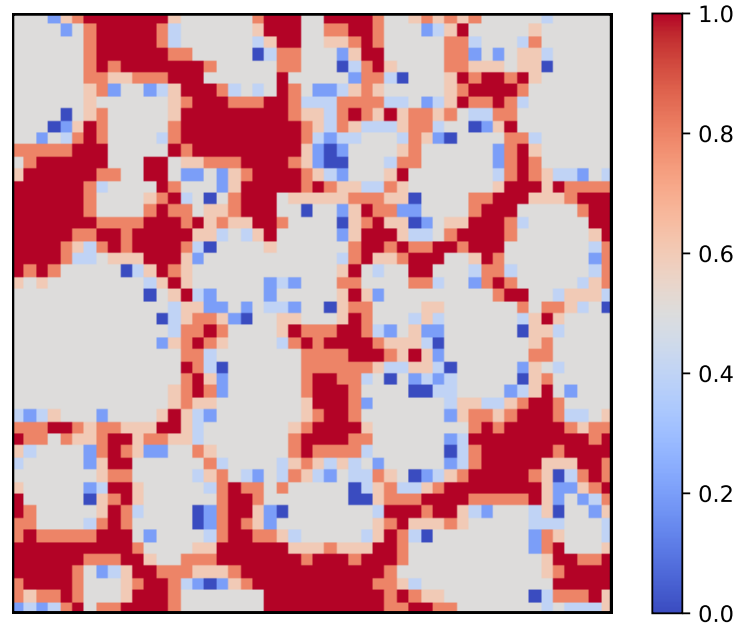}
        \caption{}
        \label{fig6:f}
    \end{subfigure}
    \hfill
    \begin{subfigure}[b]{0.24\textwidth}
        \centering
        \includegraphics[width=\textwidth]{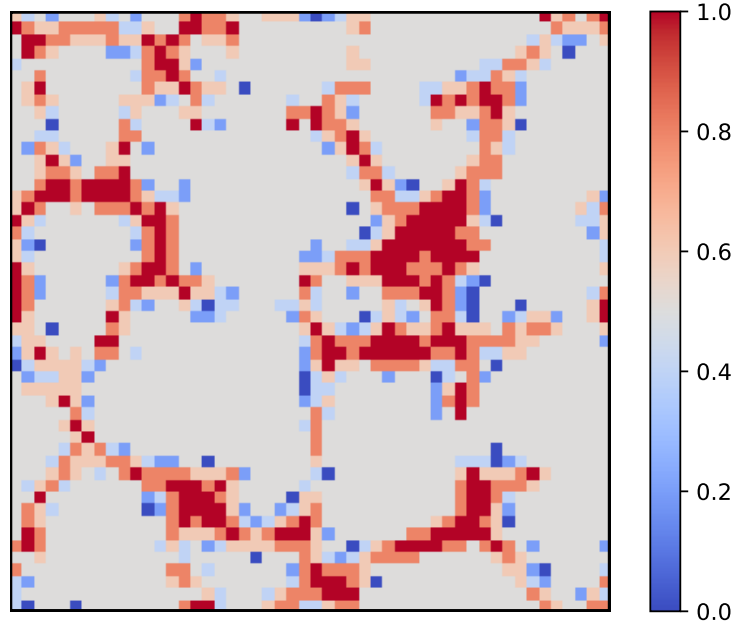}
        \caption{}
        \label{fig6:g}
    \end{subfigure}
    \hfill
    \begin{subfigure}[b]{0.24\textwidth}
        \centering
        \includegraphics[width=\textwidth]{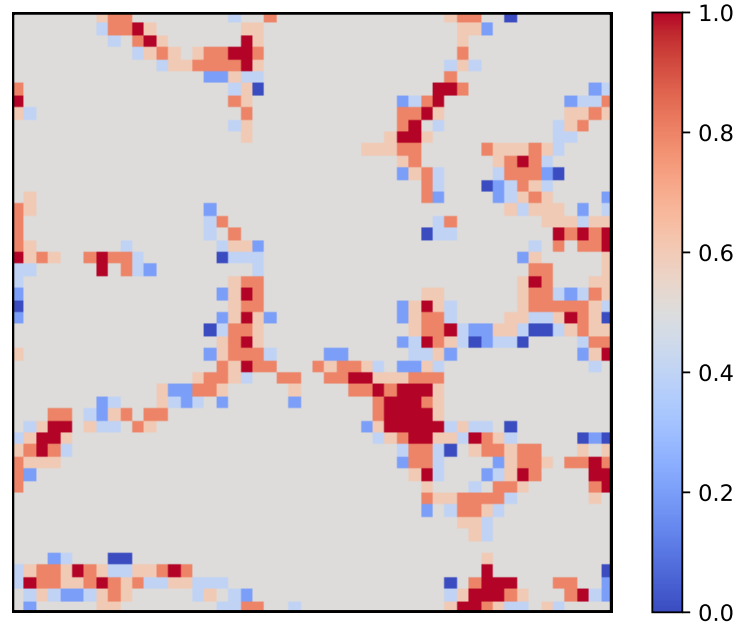}
        \caption{}
        \label{fig6:h}
    \end{subfigure}
    \caption{\textbf{Payoffs for players and referees.} Panels~(a)-(d): Payoffs for players for $r\in\{3.0,3.1,3.3,3.4\}$ in the upper row. Panels~(e)-(h): Payoffs for referees for $r\in\{3.0,3.1,3.3,3.4\}$ in the lower row. The meaning of colors is explained in the right-hand side legend for each panel. We fix $m=0.5$, $\alpha = 0.1$ and $\beta = 0.2$ for all cases. The stationary distributions are recorded after $t=3000$ Monte Carlo steps in a system containing $N=2500$ agents.}
    \label{fig6}
\end{figure*}

To complete our study we also analyze the individuals’ payoffs in the stationary state of the multilayer network. We therefore present snapshots of the payoff distribution in the game layer and the referee layer in Fig.~\ref{fig6} at different values of $r$, respectively. To ensure the stable stationary state of the network, we record the spatial distributions at $t=3000$ step. The upper panel represents the payoffs in the game layer, while the lower panel shows in the monitoring layer. The meaning of colors is explained in the right-hand side legend for each panel. It is illustrated that the overall payoffs for players gradually increase with the increase of $r$ in the game layer. It is easy to realize that cooperators gradually take advantage in the game layer as $r$ increases, and the payoffs in areas where cooperators gather into compact cluster will also be higher, which is also consistent with the results shown in Fig.~\ref{fig5}. Cooperators gather together to counteract the invasion of defectors, and the cooperation density $f_C$ increases with the increase of $r$. While the opposite situation occurs in the referee layer, the overall payoffs for referees decrease as $r$ increases. It is clear that $f_C$ and $f_F$ will increase as $r$ increases. Consequently, there are more corrupt referees in the monitoring layer when $r$ is small. Since we set $\beta=0.2$, the maximum payoff the corrupt referee can achieve from a fully defector group is 1. From Fig.~\ref{fig6}, we can also discover that the high-payoff segment in the referee layer overlaps with the low-payoff in the game layer. At this point, most of the high-payoff individuals in the referee layer are corrupt referees, whose payoffs come from defectors’ bribes. Defectors not only fail to gain additional payoffs from the game, but they also need to bribe corrupt referees, which results in lower payoffs for defectors. As $r$ increases, there are fewer defectors in the game layer and corrupt referees only receive fewer bribes, leading to a gradual increase in fair referees. Since the payoffs for fair referees originated from the fixed supervising fee $m$, hence the average payoff for referees approaches this value when fair referees dominate the referee layer. There are also referees with empty handed and earning zero individual income. They fall between two stools, neither receive bribes when facing with a group of cooperators in the game layer, not collecting the supervising fee for fair behavior.

\section{Conclusion and Outlook}\label{sec4}

In this article, we propose an interdependent network approach to model supervised cooperation in a spatially structured public goods game. The key element of our model is a realistic assumption, namely referees who monitor the actors playing the social game are not necessarily fair reviewers, but may behave improperly. Defectors in the game layer can utilize it and offer them a bribe to avoid punishment which would be otherwise a clear consequence of meeting with a fair reviewer. The other critical assumption of our approach is not only actors who play the social dilemma game can learn from each other via an imitation process, but also referees in the monitoring layer may change their attitude in dependence of their success. In this way, we can observe a coevolutionary process in the interconnected layers.
Our simulation highlighted that both the punishment ratio $\alpha$ and the $\beta$ bribe cost of defectors can be a clue of cooperation. The greater their values, the more cooperators and fair referees are present in the interdependent network. Moreover, there is a rapid phase transition between the pure cooperation and pure defection states in the game layer as the value of $\alpha$ increases, as well as between the pure fairness and pure corruption states in the referee layer. On the contrary, the consequence of $\beta$ exhibits the opposite behavior. The smaller the value of $\beta$, the sharper the phase transition. The spatial structure of fair referees shows the same clustering patterns as cooperators do, which defends the invasion of defectors and corrupt referees. The supervision fee $m$ also has a significant effect on the emergence of fair referees. If $m$ is small enough, the payoffs for fair referees are lower than the average income of corrupt partners, which will result in corrupt referees prevail on the entire referee layer. Our important observation was to realize the simultaneous and correlated pattern formation in both layers, which is a direct consequence of so-called interdependent network reciprocity frequently reported in the applied topology. 

We also analyzed the stationary payoff distribution in the game layer and in the referee layer. As synergy factor $r$ increases, the average payoff in the game layer gradually rises. In the referee layer, most of the referees’ payoffs are from the supervising fee earned by fair referees. Our results are consistent with the real-life situations, where appropriately increasing the penalties for illegal behavior can effectively reduce its occurrence. Additionally, raising the income of government officials can effectively decrease corruption, thus promoting a harmonious society.

This work can be extended and further studied from the following perspectives. The punishment fines and bribes can be a nonlinear function. The cooperation emergence of nonlinear selection due to the complex bribe structure is also an interesting research path. Notably, our model is just the very first step because it assumes that structure of game layer and referee layers are identical for simplicity. The effect of different game interaction and referee networks can be further taken into account. The effect of fair and corrupt referees on higher-order networks may also result in more subtle system behavior which can extend our present understanding. 

\section*{Acknowledgment}

This work is supported by the Natural Science Foundation of Chongqing (Grant No.CSTB2023NSCQ-MSX0064) and the National Natural Science Foundation of China (NSFC) (Grant No.62206230), and the National Research, Development and Innovation Office (NKFIH) under Grant No. K142948.

\printcredits

\bibliographystyle{model1-num-names}

\bibliography{citation}
\end{document}